\documentclass[copyright,creativecommons]{eptcs} 
\providecommand{\event}{FORECAST 2016} 
\usepackage{breakurl}             
\usepackage{graphicx}
 
\usepackage{amsmath,amssymb,latexsym}
\usepackage{pgf}
\usepackage{tikz}
\usetikzlibrary{arrows,automata}

\usetikzlibrary{shapes.symbols}

\usepackage{carmamath}
\definecolor{ForestGreen}{rgb}{0.158, 0.988, 0.178}

\title{Data as processes:\\ introducing measurement data into CARMA models}
\author{Stephen Gilmore
\institute{Laboratory for Foundations of Computer Science\\
The University of Edinburgh\\
Edinburgh, Scotland}
\email{Stephen.Gilmore@ed.ac.uk}
}
\def\titlerunning{Data as processes}
\def\authorrunning{S. Gilmore}

\begin{document}
\maketitle

\begin{abstract}
Measurement data provides a precise and detailed description of components within a complex system but it is rarely used directly as a component of a system model.  In this paper we introduce a model-based representation of measurement data and use it together with modeller-defined components expressed in the CARMA modelling language.  We assess both liveness and safety properties of these models with embedded data.
\end{abstract}

\section{Introduction}

A formal model of a real-world system uses abstraction to distill the most important elements of the system into a succinct representation which is amenable to formal reasoning and analysis.  If the modeller creating the model has chosen the right level of abstraction for their analysis then the insights which are gained from model-based reasoning are also applicable to the real-world system itself.  If, however, some of the important elements of the system have been mis-represented in the model then the insights gained by model-based reasoning and analysis are of no value, no matter how much trouble or care was taken to obtain them from the (flawed) formal model.

For dynamic models used to study performance properties such as throughput, utilisation, and satisfaction of service-level agreements, one challenge which the modeller must face is representing the timed behaviour of systems accurately.  Depending on the kind of model that is being created, either continuous sure variables or random variables from a particular random number distribution are used to abstract aspects of timed behaviour in the system under study.  These variables are parameters of the model, allowing it to be used in a suite of experiments which explore the behaviour of the model when some of the parameters are perturbed. For such a modelling study to be informative about the system under study it is then necessary to ensure that these parameters are correctly chosen to reflect the durations of the corresponding events in the system.  

Techniques for abstracting empirical univariate distributions into statistical distributions such as phase-type distributions are well known and available as algorithms~\cite{10.2307/4616418} and even as software tools~\cite{DBLP:conf/epew/ReineckeKW13}. However, in the case of systems where spatial aspects play a significant role in addition to timed behaviour we have several correlated variables and a multivariate distribution which means that finding a suitable abstraction is not so easy.
Where a classical model of a spatially-distributed system would typically use a co-ordinate system to provide an abstract representation of space, our concrete component instead uses literal latitude and longitude co-ordinates to represent the current position of a mobile component. The result is a model which is a mix of abstract components crafted by the modeller and concrete components which have been automatically generated from measurement data.  This allows us to build models of systems where we selectively choose not to abstract one component, but instead to represent it literally in order to ensure that we do not misrepresent it via an inappropriate abstraction.

From the viewpoint of model-based testing we should see each concrete component as a \emph{black box} component within the model.  The component offers up the values of its attributes at any time, but the logic as to why the attribute values change as they do is not represented anywhere in the model, neither in the concrete components generated from measurement data nor in the abstract components defined by the modeller.  Measurement data can be easily obtained from an instrumented system and one can often be in the situation of having an embarrassingly large volume of measurement data.  Because the concrete components admit no compact representation of their behaviour the modelling formalism which we use must be able to tolerate large unstructured components with real-valued attributes such as latitude and longitude, paired with timestamps.

Many modelling formalisms are not able to meet this challenge.  Classical Petri nets, process algebras, and layered queueing networks do not provide the data types and data structures which are needed to represent concrete components within the model.  Here however, we are working with CARMA (Collective Adaptive Resource-sharing Markovian Agents)~\cite{DBLP:journals/corr/BortolussiNGGHL15} a modern feature-rich modelling language which in addition to providing a stochastic process algebra of guarded recursive processes with unicast and broadcast communication also provides the primitive data types and data structures of a general-purpose programming language.  These features are supplemented by encapsulation mechanisms, general function definitions, and iterative constructs for defining collectives of components.  Together these features give the modeller sufficient linguistic power to represent concrete components directly within CARMA models, and we utilise this strength of CARMA modelling here.

\section{Background}

Models in the CARMA language consist of a \emph{collective} of \emph{components}, set in an \emph{environment} representing the context in which the components operate.  The collective is a parallel composition of components, each of which consists of a \emph{process} which represents the component's behaviour, and a \emph{store} which represents the component's knowledge.

Stores map \emph{attribute names} to \emph{basic values} of primitive types such as boolean, integer and real.  Values such as these can be passed as parameters when processes communicate.  An output action $\alpha\langle\vec{v}\rangle$ by one process can be matched with an input action $\alpha(\vec{x})$ by another process provided the length of the vector of values $\vec{v}$ is the same as the length of the vector of variables $\vec{x}$.  The arity of a communication action must be consistent throughout the model: it cannot be used to pass one value at one point and two (or more) values at another. Communication actions can either be \emph{unicast} or \emph{broadcast} in CARMA\@.  In all, this provides four types of actions in CARMA:
\begin{displaymath}
\begin{array}{rlp{295pt}}
\hbox{broadcast output} 
	& \alpha^\star[\pi]\langle\vec{e}\rangle\sigma
	& asynchronous (non-blocking) broadcast action $\alpha$ to communication partners identified by the predicate $\pi$; send the values of expressions $\vec{e}$ evaluated in the local store $\gamma$; then apply the update $\sigma$ to $\gamma$.
\\
\hbox{broadcast input} 
	& \alpha^\star[\pi](\vec{x})\sigma
	& receive a tuple $\vec{x}$ of values $\vec{v}$ sent with an action $\alpha$ from a component whose store satisfies the predicate $\pi[\vec{v}/\vec{x}]$; then apply the  update $\sigma$ to local store $\gamma$.
\\
\hbox{unicast output} 
	& \alpha[\pi]\langle\vec{e}\rangle\sigma
	& synchronous (blocking) unicast action $\alpha$ to any communication partner  satisfying the predicate $\pi$; send the values of expressions $\vec{e}$ evaluated in the local store $\gamma$; then apply the update $\sigma$ to $\gamma$.
\\
\hbox{unicast input} 
	& \alpha[\pi](\vec{x})\sigma
	& (point-to-point) receive a tuple $\vec{x}$ of values $\vec{v}$ sent with an action $\alpha$ from a component whose store satisfies the predicate $\pi[\vec{v}/\vec{x}]$; then apply the update $\sigma$ to local store $\gamma$.
\\
\end{array}
\end{displaymath}
The use of predicates to describe communication partners means that CARMA supports the \emph{attribute-based communication} paradigm~\cite{DBLP:conf/sac/AlrahmanNLTV15}, as found in languages such as SCEL~\cite{DBLP:series/lncs/NicolaLLLMMMPTV15}, where dynamic collections of components called \emph{ensembles} are formed through having attributes in common.  In process algebras such as PEPA~\cite{pepa} where data is abstracted out of the model, attributes are not present and thus attribute-based communication is not possible.  Communication partners are determined statically in PEPA whereas they are determined dynamically in CARMA and SCEL.

\newcommand{\blueDef}{\textcolor{blue}{::=}}
\newcommand{\blueOr}{\mathrel{\textcolor{blue}{\mid}}}
Processes ($P, Q$, \ldots) in CARMA are defined by the following grammar:
\begin{displaymath}
\begin{array}{rcl}
P, Q & \blueDef & \mathbf{nil} 
       \blueOr  \mathbf{kill}
       \blueOr  \mathit{act}.P
       \blueOr P+Q 
       \blueOr {P\vert Q}
       \blueOr [\pi]P
       \blueOr A \ \  (A \triangleq P)
\\
\mathit{act} & \blueDef &  \alpha^\star[\pi]\langle\vec{e}\rangle\sigma
       \blueOr \alpha^\star[\pi](\vec{x})\sigma
       \blueOr \alpha[\pi]\langle\vec{e}\rangle\sigma
       \blueOr \alpha[\pi](\vec{x})\sigma
\end{array}
\end{displaymath}
By convention in a CARMA model activity names begin with a lowercase letter, function and component names begin with a capital letter, and process names are written in all caps.  Expressions in the CARMA language (as used in function bodies) are generated by the following grammar.
\begin{displaymath}
\begin{array}{rcl}
e_1, e_2, e_3 & \blueDef & 
            \mathbf{return}~e_1 
       \blueOr \mathbf{if} (e_1) \{ e_2 \}
       \blueOr \mathbf{if} (e_1) \{ e_2 \} \mathrel{\mathbf{else}} \{ e_3 \}
       \blueOr e_1 ; e_2
       \blueOr a_1
       \blueOr b_1
\\
a_1, a_2 & \blueDef & 0 
       \blueOr 1 
       \blueOr \cdots
       \blueOr -a_1
       \blueOr a_1 + a_2
       \blueOr a_1 - a_2
       \blueOr a_1 * a_2 
       \blueOr a_1 / a_2 
\\
b_1, b_2 & \blueDef & \mathbf{true}
       \blueOr \mathbf{false}
       \blueOr a_1 > a_2
       \blueOr a_1 >= a_2
       \blueOr a_1 == a_2 
       \blueOr a_1 <= a_2
       \blueOr a_1 < a_2\\
		&\blueOr& \,{!b_1}
       \blueOr \, b_1 \mathrel{\&\&} b_2 
       \blueOr \, b_1 {\vert\vert} b_2
\\
\end{array}
\end{displaymath}

\section{Case study: Bus fleet management}

For our case study in this paper we show how data on the movement of a bus travelling through the city of Edinburgh can be incorporated into a CARMA model.  The purpose of the modelling will be to check whether or not the bus follows the intended route by matching its movements against a high-level description of the route in terms of regions of the city described by predicates.  This is an instance of a \emph{fleet management} problem as studied in the transportation modelling community: it is important to know the location of all of the vehicles in the fleet and to know that they are serving their assigned routes.  Managing the assignment of buses to routes is not as easy in practice as it might appear: changes of assignment are needed during the working day as problems such as vehicular mechanical failures, road closures, or driver unavailabilty can cause buses to be cancelled or re-routed in ways that would be impossible to predict at the start of the day.

\begin{figure}[htbp]
\centering{\includegraphics[width=0.8\textwidth,trim=50 50 30 300,clip]{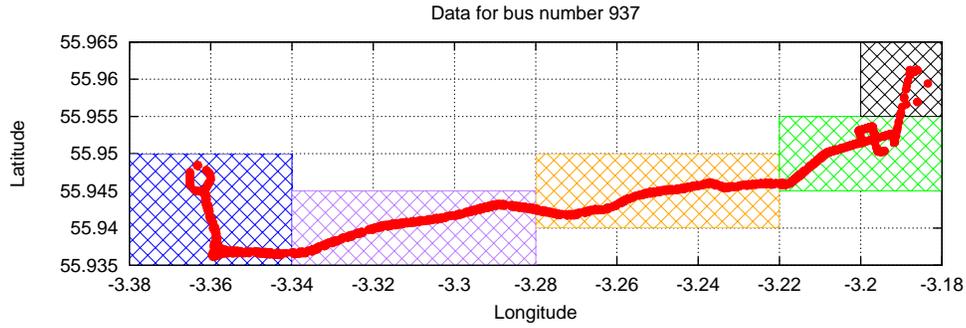}}
\caption{\label{fig:plot.937.five.regions} The route of bus number 937 in the fleet, assigned to service 100. The route includes the airport (in the bottom left hand corner, in blue), suburban area~1 (in purple), suburban area~2 (in orange), the city centre (in green), and the garage (in the top right hand corner, in black).}
\end{figure}

Our specific example is Transport for Edinburgh's Service 100, which travels between Edinburgh airport and Edinburgh city centre.  We can characterise this route as having five significant regions: the airport, suburban area~1, suburban area~2, the city centre, and the garage where the bus is parked overnight.  These areas are shown in Figure~\ref{fig:plot.937.five.regions}, together with a GPS trace of bus fleet number~937 serving this route.  The definition of these regions is given in Table~\ref{tab:table.of.region.definitions}. 

\newcommand{\Lat}[1]{\mathit{lat}_{#1}}
\newcommand{\Long}[1]{\mathit{long}_{#1}}
\newcommand{\Airport}{\mathit{airport}}
\newcommand{\Region}[1]{\big[ #1 \big]}
\newcommand{\Suburbs}[1]{\mathit{suburbs}_{#1}}
\newcommand{\Centre}{\mathit{centre}}
\newcommand{\Garage}{\mathit{garage}}

\begin{table}[htbp]
\begin{displaymath}
\begin{array}{rcl c|c rcl c|c rcl}
\multicolumn{4}{c|}{\mathit{Significant}} & 
\multicolumn{5}{c|}{\mathit{Significant}} & 
\multicolumn{4}{c}{\mathit{Region}}\\
\multicolumn{4}{c|}{\mathit{latitudes}} & 
\multicolumn{5}{c|}{\mathit{longitudes}} & 
\multicolumn{4}{c}{\mathit{definitions}}\\[2pt]
\hline
          &  &                 &&&             & &           &&&              &  &  \\[-6pt]
\Lat1 &=& 55.935  	&&& \Long1 &=& -3.38 &&& \Airport &=& \Region{(\Long1, \Lat1) , (\Long2, \Lat4)}\\[2pt]
\Lat2 &=& 55.940	&&& \Long2 &=& -3.34 &&& \Suburbs1 &=& \Region{ (\Long2, \Lat1),  (\Long3, \Lat3) } \\[2pt]
\Lat3 &=& 55.945 	&&& \Long3 &=& -3.28 &&& \Suburbs2 &=& \Region{ (\Long3, \Lat2), (\Long4, \Lat4) } \\[2pt]
\Lat4 &=& 55.950	&&& \Long4 &=& -3.22 &&& \Centre &=& \Region{ (\Long4, \Lat3), (\Long6, \Lat5) } \\[2pt]
\Lat5 &=& 55.955      &&& \Long5 &=& -3.20 &&& \Garage &=& \Region{ (\Long5, \Lat5), (\Long6,\Lat6) }\\[2pt]
\Lat6 &=& 55.965      &&& \Long6 &=& -3.18 &&&               &   &      \\ 
\end{array}
\end{displaymath}
\caption{\label{tab:table.of.region.definitions} Table of region definitions in terms of latitude and longitude coordinates.}
\end{table}

In the dataset that we are working with here, the position of a bus has been registered once every minute.  Regions should be chosen to be large enough to make it effectively improbable that  a bus can enter the region and exit from it again without having been observed at least once within it.  Regions should bound a portion of the bus route, but not so tightly that small measurement errors in GPS readings could cause a bus to be perceived as outside that region.  We have chosen our regions to be simple rectangles because it is easy to test whether a point lies within a simple geometric shape such as this.  We defined five CARMA predicates to test whether a point lies in a region.  These are \texttt{AtAirport}, \texttt{InSuburbs1}, \texttt{InSuburbs2}, \texttt{InCentre} and \texttt{AtGarage}.  The CARMA predicate \texttt{AtAirport} is shown in Figure~\ref{fig:AtAirport.function.in.CARMA}.  The other predicates are similarly easy to define.

\begin{figure}[htbp]
\begin{lstlisting}
fun bool AtAirport(real long, real lat) {
  if (long > long1 && long < long2 && lat > lat1 && lat < lat4) {
    return true;
  } else {
    return false;
  }
}
\end{lstlisting}
\caption{\label{fig:AtAirport.function.in.CARMA} The \texttt{AtAirport} function in CARMA.}
\end{figure}

\subsection{Generating a concrete component from measurement data}

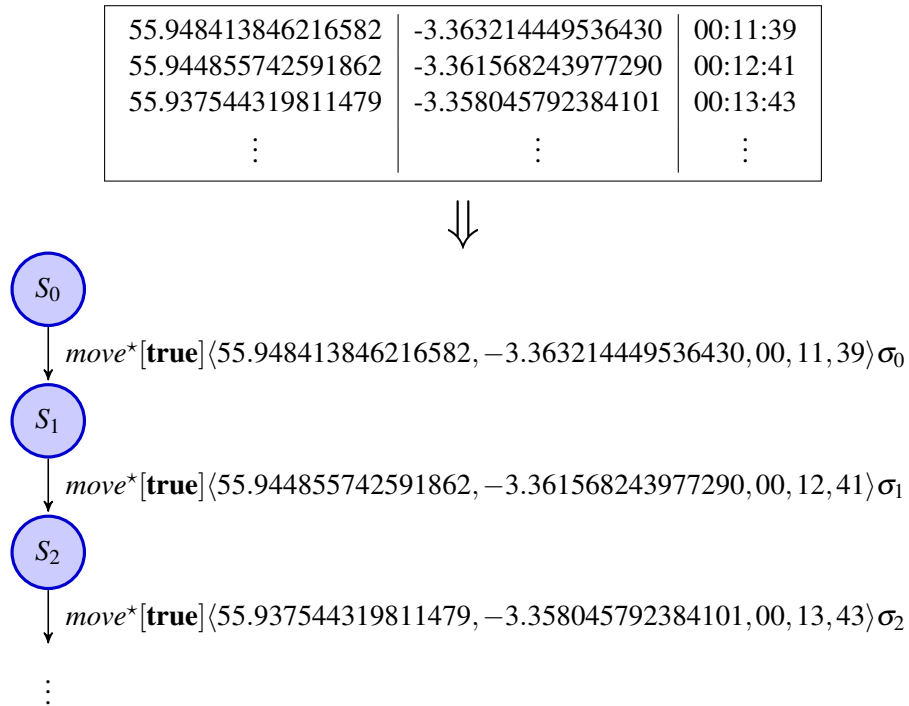
\begin{figure}[htbp]%
\begin{center}%
\framebox{\begin{tabular}{c|c|c}
55.948413846216582&-3.363214449536430&00:11:39\\
55.944855742591862&-3.361568243977290&00:12:41\\
55.937544319811479&-3.358045792384101&00:13:43\\
  \vdots & \vdots & \vdots 
\end{tabular}}

\medskip
{\huge$\Downarrow$}

\begin{tikzpicture}[->,>=stealth',shorten >=1pt,auto,node distance=1.75cm,
                    semithick]
  \tikzstyle{every state}=[fill=white,draw=none,text=black]

  \node[state,draw=blue!80!black!100,very thick,fill=blue!20] 	  (S0)                    {$S_0$};
  \node[state,draw=blue!80!black!100,very thick,fill=blue!20] 	  (S1) [below of=S0]                    {$S_1$};
  \node[state,draw=blue!80!black!100,very thick,fill=blue!20] 	  (S2) [below of=S1]                    {$S_2$};
  \node[state,draw=white,very thick,fill=white] 	  (S3) [below of=S2]                    {$\vdots$};

  \path 
      (S0) 
  		edge [solid]            							node {\ $\mathit{move^{\star}}[\textbf{true}]\langle
55.948413846216582,-3.363214449536430,00,11,39
                                                        \rangle\sigma_0$} (S1)
      (S1) 
  		edge [solid]            							node {\ $\mathit{move^{\star}}[\textbf{true}]\langle
55.944855742591862,-3.361568243977290,00,12,41
                                                        \rangle\sigma_1$} (S2)
      (S2) 
  		edge [solid]            							node {\ $\mathit{move^{\star}}[\textbf{true}]\langle
55.937544319811479,-3.358045792384101,00,13,43
                                                        \rangle\sigma_2$} (S3)
;
\end{tikzpicture}
\end{center}
\caption{\label{fig:converting}Converting measurement data into concrete components in our CARMA model.}
\end{figure}

Given measurement data whose records consist of latitude and longitude coordinates together with a timestamp it is straightforward to generate a concrete component which introduces this measurement data into our CARMA model.  We generate a straight-line process which broadcasts each \emph{move} action as it occurs.  We choose broadcast output because in general we do not want the abstract components of the model to alter the behaviour of the concrete components.  The parameters of the \emph{move} action are the measurement data in the form of a five-tuple $\langle\mathit{latitude}$, $\mathit{longitude}$, $\mathit{hour}$, $\mathit{minutes}$, $\mathit{seconds}\rangle$.  This conversion from measurement data into a CARMA component is performed automatically with a Python script.  Figure~\ref{fig:converting} illustrates this process.  The updates~$\sigma_0$, $\sigma_1$ and~$\sigma_2$ are the obvious updates of the local store to hold the current values of latitude, longitude, hours, minutes, and seconds.

\begin{figure}[htbp]
\centering{\includegraphics[width=\textwidth,trim=15 45 0 0,clip]{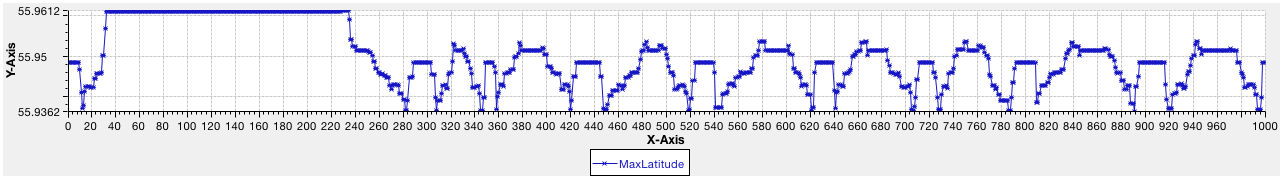}}
\caption{\label{fig:max.latitude.plot} A plot generated by the CARMA Eclipse Plugin showing the maximum value of the \emph{latitude} attribute of the \emph{Bus} component in our model.  The long constant high value of latitude near the start of the trace records the bus being parked in the garage overnight.}
\end{figure}

Now that we have our measurement data within our CARMA model we can use the CARMA Eclipse Plugin to execute the model and investigate its behaviour using a \emph{measure} defined in the CARMA model.  Measures are real-valued functions which compute some result of the current state of the model, allowing this to be visualised as an assessment of the model's behaviour.  We can define a measure in CARMA as shown below.
\begin{lstlisting}
    measure MaxLatitude = max { my.latitude };
\end{lstlisting}
The \emph{Bus} component which we have generated has an attribute in its local state named \emph{latitude} which is updated after every movement action.  A component refers to its own local state using the prefix \textbf{my} in CARMA (much like the use of \textbf{this} in Java). The particular measure shown above records the maximum value of the latitude seen at all timepoints along the trace of the \emph{Bus} component as it executes.  The results are shown in Figure~\ref{fig:max.latitude.plot}.

This plot assures us that the model exhibits \emph{some} behaviour, in that the latitude of the bus is changing, but we do not yet know whether or not it is following the route of the~100 service, or staying within the five regions of interest specified earlier.  We will define an additional component to investigate these questions.

\subsection{Defining a probe to monitor the concrete component}

The next step in checking the correctness of a bus journey is to be able to monitor its behaviour by adding an abstract component (defined by the modeller) to ensure that the progress of the bus moving between regions is as we expect, and that the bus does not leave the five regions which we have defined (see Table~\ref{tab:table.of.region.definitions}).  

We use the term \emph{probe} for a component whose purpose is simply to monitor changes in another component~\cite{stochasticprobes,xsp}.  The function of a probe is to make it convenient to express checkable properties of a model.  A probe is a finite-state automaton which recognises the language of acceptable state transitions within a model and rejects all unacceptable state transition sequences.  Probes are sometimes expressed in formal language terms as regular expressions and sometimes as timed automata~\cite{DBLP:conf/dsn/AmparoreD10,DBLP:conf/qest/AmparoreBDF11}.

Probes can be used to check both safety and liveness properties of models.  Here we are interested in two liveness properties and one safety property, as described below.
\begin{description}
\item[Liveness 1:] The bus visits the airport (probe reaches state \texttt{AIRPORT}).
\item[Liveness 2:] The bus visits the city centre (probe reaches state \texttt{CENTRE}).
\item[Safety:] The bus does not leave the five defined regions (probe never reaches state \texttt{ERROR}).
\end{description}

\begin{figure}[htbp]
\begin{center}
\begin{tikzpicture}[->,>=stealth',shorten >=1pt,auto,node distance=2cm,
                    semithick]
  \tikzstyle{every state}=[fill=white,draw=none,text=black]

  \node[initial,state,draw=blue!80!black!100,very thick,fill=blue!20] 	  	(S0)                    {$A$};

  \node[state,draw=red!80!black!100,very thick,fill=red!60]            	(S5) [right of=S0] {$E$};

  \node[state,draw=black!80!black!100,very thick,fill=black!20]            	(S4) [right of=S5] {$G$};

  \node[state,draw=ForestGreen!60!black!100,very thick,fill=ForestGreen!20]                    	(S3) [right of=S4] {$C$};

  \node[state,draw=purple!80!orange!100,very thick,fill=purple!20]        	(S1A) [above of=S5] {$S_1^A$};
  \node[state,draw=orange!80!black!100,very thick,fill=orange!20]        	(S2A) [above of=S4] {$S_2^A$};

  \node[state,draw=purple!80!orange!100,very thick,fill=purple!20]        	(S1B) [below of=S5] {$S_1^R$};
  \node[state,draw=orange!80!black!100,very thick,fill=orange!20]        	(S2B) [below of=S4] {$S_2^R$};

  \path 
      (S0) 
  		edge [solid]            										node {} (S1A)
      (S0) 
  		edge [solid]            										node {} (S5)
      (S0) 
  		edge [loop below,solid]            							node {} (S0)

      (S1A) 
  		edge [solid]            										node {} (S2A)
      (S1A) 
  		edge [solid]            										node {} (S5)
      (S1A) 
  		edge [loop above,solid]            							node {} (S1A)

      (S2A) 
  		edge [solid]            										node {} (S3)
      (S2A) 
  		edge [bend right=0,solid]            						node {} (S5)
      (S2A) 
  		edge [loop above,solid]            							node {} (S2A)

      (S3) 
  		edge [bend right=20,solid]            						node {} (S4)
      (S3) 
  		edge [solid]            										node {} (S2B)
      (S3) 
  		edge [bend right=30,solid]            						node {} (S5)
      (S3) 
  		edge [loop below,solid]            							node {} (S3)

      (S4) 
  		edge [bend right=20,solid]            						node {} (S3)
      (S4) 
  		edge [solid]            										node {} (S5)
      (S4) 
  		edge [loop below,solid]            							node {} (S4)

      (S2B) 
  		edge [bend left=0,solid]            						node {} (S5)
      (S2B) 
  		edge [solid]            										node {} (S1B)
      (S2B) 
  		edge [loop below,solid]            							node {} (S2B)

      (S1B) 
  		edge [solid]            										node {} (S0)
      (S1B) 
  		edge [bend left=40,solid]            						node {} (S5)
      (S1B) 
  		edge [loop below,solid]            							node {} (S1B)

      (S5) 
  		edge [loop below,solid]            							node {} (S5)
;
\end{tikzpicture}
\end{center}
\caption{\label{fig:probe.Allez.Retour}A probe component which separates the outward and return routes.}
\end{figure}
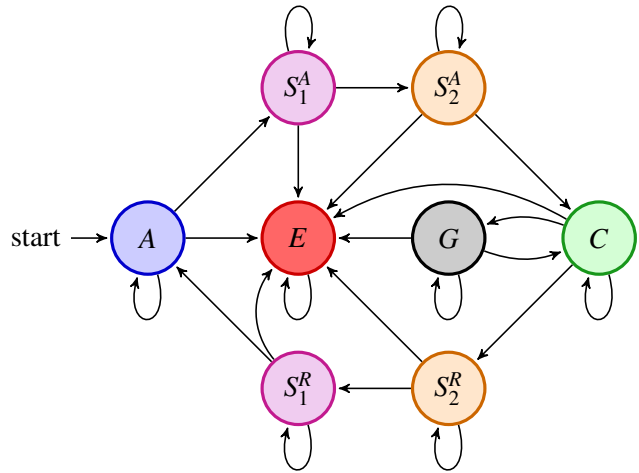

Perhaps the most natural description of the journey of the bus would be to separate out the Airport journey (from the airport to the city centre) and the Return journey (from the city centre to the airport).  Denoting these $A$ and $R$ respectively, we would note that the journey from the airport to the city centre passes through the two suburban regions in the order $[S_1^A; S_2^A]$ whereas the return journey passes through the two suburban regions in the order $[S_2^R; S_1^R]$\@.  The probe which captures this separation of the Airport and Return journeys is presented in Figure~\ref{fig:probe.Allez.Retour}.  The predicate guards which label each arrow have been omitted to reduce clutter.  The selection of start state means that this probe can only be applied to vehicles which begin their journey at the airport.  State $E$ indicates that an error has occurred.

Although this description of the probe is perfectly correct from the abstract notion of the bus route it is in practice rather too unforgiving of measurement errors.  For a bus stopped on the border between the region  $\Suburbs1$ and the region $\Suburbs2$ a small error in GPS measurement could cause the sequence of observations $[S_1^A; S_2^A; S_1^A; S_2^A]$ to be seen, and this cannot be accepted by the probe presented in Figure~\ref{fig:probe.Allez.Retour}.

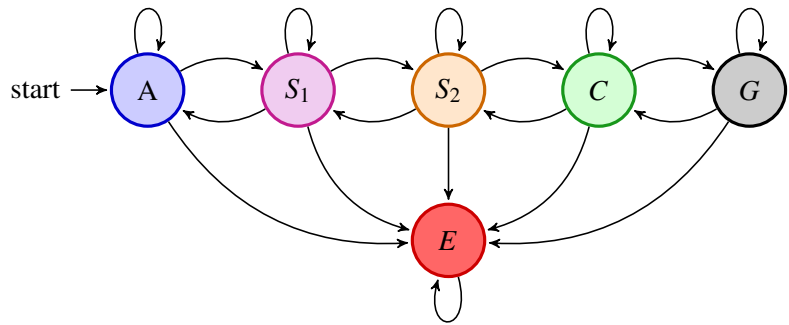
\begin{figure}[!tbh]
\begin{center}
\begin{tikzpicture}[->,>=stealth',shorten >=1pt,auto,node distance=2cm,
                    semithick]
  \tikzstyle{every state}=[fill=white,draw=none,text=black]

  \node[initial,state,draw=blue!80!black!100,very thick,fill=blue!20] 	  (S0)                    {A};
  \node[state,draw=purple!80!orange!100,very thick,fill=purple!20]        (S1) [right of=S0] {$S_1$};
  \node[state,draw=orange!80!black!100,very thick,fill=orange!20]        (S2) [right of=S1] {$S_2$};
  \node[state,draw=ForestGreen!60!black!100,very thick,fill=ForestGreen!20]                    (S3) [right of=S2] {$C$};
  \node[state,draw=black!80!black!100,very thick,fill=black!20]              (S4) [right of=S3] {$G$};
  \node[state,draw=red!80!black!100,very thick,fill=red!60]            (S5) [below of=S2] {$E$};

  \path 
      (S0) 
  		edge [bend left=30,solid]            							node {} (S1)
      (S0) 
  		edge [loop above,solid]            								node {} (S0)

      (S1) 
  		edge [bend left=30,solid]            							node {} (S0)
      (S1) 
  		edge [loop above,solid]            								node {} (S1)

      (S1) 
  		edge [bend left=30,solid]            							node {} (S2)

      (S2) 
  		edge [bend left=30,solid]            							node {} (S1)
      (S2) 
  		edge [bend left=30,solid]            							node {} (S3)
      (S2) 
  		edge [loop above,solid]            								node {} (S2)

      (S3) 
  		edge [bend left=30,solid]            							node {} (S2)
      (S3) 
  		edge [bend left=30,solid]            							node {} (S4)
      (S3) 
  		edge [loop above,solid]            								node {} (S3)

      (S4) 
  		edge [bend left=30,solid]            							node {} (S3)
      (S4) 
  		edge [loop above,solid]            								node {} (S4)

      (S0) 
  		edge [bend right=30,solid]            							node {} (S5)
      (S1) 
  		edge [bend right=30,solid]            							node {} (S5)
      (S2) 
  		edge [solid]            											node {} (S5)
      (S3) 
  		edge [bend left=30,solid]            							node {} (S5)
      (S4) 
  		edge [bend left=30,solid]            							node {} (S5)
      (S5) 
  		edge [loop below,solid]            								node {} (S5)
;
\end{tikzpicture}
\end{center}
\caption{\label{fig:probe.loose.specification}A probe component which does not differentiate between the outward and return routes.}
\end{figure}

For this reason, we work with a looser specification of the bus route, as described by the probe in Figure~\ref{fig:probe.loose.specification}.  This does not differentiate between the Airport and Return routes and has the dual benefits of being more compact and tolerating errors in GPS measurement at the boundaries between regions.  For example, the sequence of observations $[S_1; S_2; S_1; S_2]$ can be accepted by this component.  The predicate guards which label each arrow have again been omitted in this diagram to reduce clutter.

Of course, when this probe is expressed as a CARMA component it is necessary to be absolutely specific about the predicate guards on each transition.  The overall effect of the predicates is to track the location of the bus on the basis of its reported latitude and longitude.  A transition to the ERROR state of the probe may only be taken if no other outgoing transition from a state is possible.  We use the predicates \texttt{AtAirport}, \texttt{InSuburbs1}, \texttt{InSuburbs2}, \texttt{InCentre} and \texttt{AtGarage} as described previously.  The text of the probe as a CARMA component is shown in Figure~\ref{fig:The.probe.as.a.CARMA.component}.

\begin{figure}[!t]
\begin{small}
\begin{lstlisting}
component Probe(process Z) {
  store {}

  behaviour {

    AIRPORT  = move%\CARMAstar%[AtAirport(long,lat)](lat,long,h,m,s){}.AIRPORT
              + move%\CARMAstar%[InSuburbs1(long,lat)](lat,long,h,m,s){}.SUBURBS1
              + move%\CARMAstar%[!AtAirport(long,lat) &&
                      !InSuburbs1(long,lat)](lat,long,h,m,s){}.ERROR;

    SUBURBS1 = move%\CARMAstar%[AtAirport(long,lat)](lat,long,h,m,s){}.AIRPORT
              + move%\CARMAstar%[InSuburbs1(long,lat)](lat,long,h,m,s){}.SUBURBS1
              + move%\CARMAstar%[InSuburbs2(long,lat)](lat,long,h,m,s){}.SUBURBS2
              + move%\CARMAstar%[!AtAirport(long,lat) &&
                      !InSuburbs1(long,lat) &&
                      !InSuburbs2(long,lat)](lat,long,h,m,s){}.ERROR;

    SUBURBS2 = move%\CARMAstar%[InSuburbs1(long,lat)](lat,long,h,m,s){}.SUBURBS1
              + move%\CARMAstar%[InSuburbs2(long,lat)](lat,long,h,m,s){}.SUBURBS2
              + move%\CARMAstar%[InCentre(long,lat)](lat,long,h,m,s){}.CENTRE
              + move%\CARMAstar%[!InSuburbs1(long,lat) &&
                      !InSuburbs2(long,lat) &&
                      !InCentre(long,lat)](lat,long,h,m,s){}.ERROR;

    CENTRE   = move%\CARMAstar%[InSuburbs2(long,lat)](lat,long,h,m,s){}.SUBURBS2
              + move%\CARMAstar%[InCentre(long,lat)](lat,long,h,m,s){}.CENTRE
              + move%\CARMAstar%[AtGarage(long,lat)](lat,long,h,m,s){}.GARAGE
              + move%\CARMAstar%[!InSuburbs2(long,lat) &&
                      !InCentre(long,lat) &&
                      !AtGarage(long,lat)](lat,long,h,m,s){}.ERROR;

    GARAGE   = move%\CARMAstar%[InCentre(long,lat)](lat,long,h,m,s){}.CENTRE
              + move%\CARMAstar%[AtGarage(long,lat)](lat,long,h,m,s){}.GARAGE
              + move%\CARMAstar%[!InCentre(long,lat) &&
                      !AtGarage(long,lat)](lat,long,h,m,s){}.ERROR;

    ERROR    = move%\CARMAstar%[true](lat,long,h,m,s){}.ERROR;
  }

  init { Z }
}
\end{lstlisting}
\end{small}
\caption{\label{fig:The.probe.as.a.CARMA.component} The probe from Figure~\ref{fig:probe.loose.specification}   represented as a CARMA component.}
\end{figure}

\subsection{Computing liveness and safety properties using probes}

\begin{figure}[htbp]
\centering{\includegraphics[width=\textwidth,trim=15 45 0 0,clip]{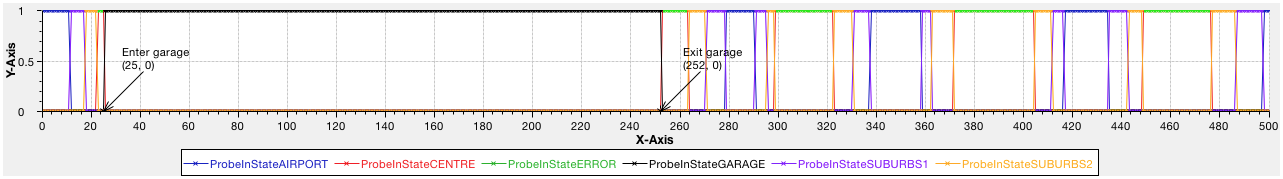}}
\caption{\label{fig:probe.937.accepted} The route of bus number 937 in the fleet. The bus is initially at the airport and enters the garage shortly after midnight.  The following day the bus performs the expected route between the city centre and the airport.  The ERROR state of the probe is never reached for bus 937.}
\end{figure}

Now we are in a position to be able to use CARMA's measures to interrogate the probe to see states which it visits.  In order to turn our probe's observations into numerical measures we count the number of probes in each state.  These measures will only ever return~0 or~1 as their results with~1 indicating that the state (\texttt{AIRPORT}, \texttt{SUBURBS1}, \ldots) has been visited.  We add these measures to our model.
\begin{lstlisting}
    measure ProbeInStateAIRPORT = #{ Probe[AIRPORT] | true };
    measure ProbeInStateSUBURBS1 = #{ Probe[SUBURBS1] | true };
    measure ProbeInStateSUBURBS2 = #{ Probe[SUBURBS2] | true };
    measure ProbeInStateCENTRE = #{ Probe[CENTRE] | true };
    measure ProbeInStateGARAGE = #{ Probe[GARAGE] | true };
    measure ProbeInStateERROR = #{ Probe[ERROR] | true };
\end{lstlisting}
There are eleven buses from the Transport for Edinburgh fleet which serve the Airport route at any time. For the day of data which we processed here, these are fleet numbers~937, 938, 939, 940, 941, 943, 945, 947, 948 and~950.  We began with bus number 937 and used the CARMA Eclipse Plugin to check our probe against the trajectory of this bus.  This showed that the two liveness properties were satisfied (the bus visits the airport and the city centre) and that the safety property was also met (the ERROR state of the probe is never reached).  The output from the CARMA Eclipse Plugin is shown in Figure~\ref{fig:probe.937.accepted}.

After this, we applied the same analysis to the remaining ten buses which were serving the~100 route, compiling these results into Table~\ref{tab:probe.100.results}.  All but one of these buses passed both the liveness tests and the safety test.  The bus which failed these tests was bus~947, which fails both of the liveness tests and also fails the safety test.

\begin{table}[htbp]
\begin{center}
\begin{small}
\begin{tabular}{l|l|l|l|l|l|l}
Fleet 	& Initial 	& Final 		& AIRPORT 	& CENTRE 	& ERROR 	& Probe 100\\
number 	& state 		& state 		& visited 	& visited 	& seen 	& result \\ 
\hline\hline
937 		& AIRPORT 	& GARAGE 	& Yes 		& Yes 		& No 		& Accepted \\
938 		& GARAGE 	& AIRPORT 	& Yes 		& Yes 		& No 		& Accepted \\
939 		& GARAGE 	& CENTRE 	& Yes 		& Yes 		& No 		& Accepted \\
940 		& SUBURBS1 	& CENTRE 	& Yes			& Yes			& No 		& Accepted \\
941 		& CENTRE 	& GARAGE 	& Yes			& Yes			& No 		& Accepted \\
943 		& GARAGE 	& GARAGE 	& Yes			& Yes			& No 		& Accepted \\
944 		& SUBURBS2 	& GARAGE		& Yes 		& Yes			& No 		& Accepted \\
945 		& CENTRE		& GARAGE 	& Yes 		& Yes			& No 		& Accepted \\
947 		& GARAGE		& ERROR		& No			& No 			& Yes		& Rejected \\
948 		& GARAGE		& AIRPORT	& Yes			& Yes			& No		& Accepted \\
950 		& GARAGE		& SUBURBS1 	& Yes			& Yes			& No		& Accepted \\
\end{tabular}
\end{small}
\end{center}
\caption{\label{tab:probe.100.results}Results of checking our probe against traces from different buses serving the 100 route.}
\end{table}

\begin{figure}[htbp]
\centering{\includegraphics[width=\textwidth,trim=15 45 0 0,clip]{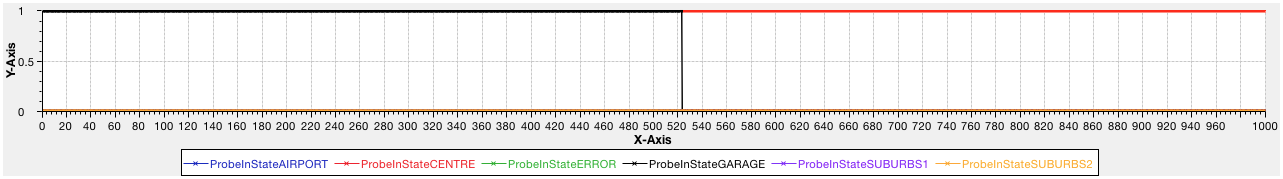}}
\caption{\label{fig:probe.947.rejected} The route of bus number 947 in the fleet. The bus is initially in the garage at the beginning of the trace but enters the ERROR state immediately on exiting the garage.}
\end{figure}

Looking at the results from the CARMA Eclipse Plugin shown in Figure~\ref{fig:probe.947.rejected}, the bus is initially in the garage (probe state is \texttt{GARAGE}) but immediately violates the allowable conditions on its latitude and longitude coordinates on leaving the garage (probe state is \texttt{ERROR}).  The error state of the probe is an absorbing state so once the probe has entered this state it will not escape to any non-error state even if the bus later corrects its position to rejoin the correct route for the service.

\begin{figure}[htbp]
\centering{\includegraphics[width=0.8\textwidth,trim=50 50 30 300,clip]{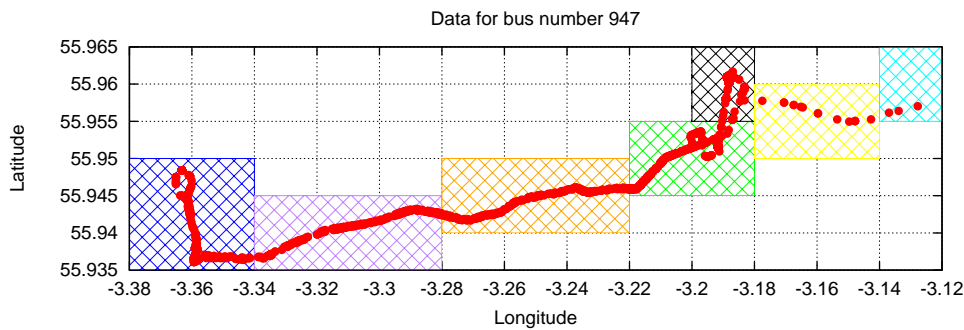}}
\caption{\label{fig:plot.947} The route of bus number 947 in the fleet, assigned to service 100. The route includes the expected locations of the airport (in the bottom left hand corner, in blue), suburban area~1 (in purple), suburban area~2 (in orange), the city centre (in green), and the garage (in black), as well as the unexpected locations of suburban area~3 (in yellow) and garage~2 (in the top right-hard corner, in cyan).}
\end{figure}

Our method of compiling measurement data into concrete components and evaluating it with a probe component has had the desired outcome of finding erroneous behaviour in an unlabelled collection of correct and incorrect trajectories.  Bus~947 has been identified as having diverged from the expected route.  We now look at its trajectory as a latitude-longitude trace and see if we can conjecture what happened to cause this deviation.  Comparing the position with a map of the city of Edinburgh we see that the bus has taken a route away from the city centre towards the coast.  A second Transport for Edinburgh garage is located here, and we can conjecture that this bus had a fault or needed some maintenance activity before it was able to serve the~100 route.  After visiting the second garage the (now, presumably, repaired) bus returns to the city centre and begins its service from there, as detailed in Figure~\ref{fig:plot.947}.

\subsection{Practicality of the method}

In our case study here we used GPS measurement data from bus journeys to build our concrete components.  The position of the bus is sampled every minute of a twenty-four hour period, thus giving us approximately~1,440 timestamped records of the latitude and longitude of the bus.  The measurement data is compiled into a CARMA model by a Python script; this CARMA model is~14,546 lines long.  This CARMA model is then compiled into a Java application by the CARMA Eclipse Plugin; this Java application is~81,162 lines long.  

The current compilation strategy employed by the CARMA compiler is to compile a CARMA component into a single Java method.  For hand-built components created by the modeller this approach has no significant disadvantages but for large generated concrete components this approach runs the risk of overflowing the maximum Java method size of~65,534 bytes.  If working with very large measurement data sets, over longer time periods or with finer sampling granularity, then it would be necessary to change the CARMA compilation strategy to compile CARMA processes into individual methods instead and have the compilation images of CARMA components call these methods, thereby moving the problem of maximum method size to reappear again only for CARMA processes which have many attributes and very large update blocks.

\section{Related work}

In this work we have considered using data concretely as process components in a model.  In earlier work, van der Aalst \textsl{et al} have devised algorithms for extracting compact process descriptions from data such as system event logs.  These algorithms are necessarily incomplete and cannot always find a compact process representation which faithfully encodes an expansive event log.  Nonetheless, these \emph{workflow mining}~\cite{van2004workflow} and \emph{process mining}~\cite{ProcessMining} approaches give valuable insights into large data sets by identifying likely causal relations between events and variants of the $\alpha$ algorithm which underlies the workflow mining approach are able to rediscover large classes of processes from event logs.

The \emph{Traviando} simulation trace analyser can also be applied to inverse problems like these, in that it can be used to generate so-called \emph{likely invariants} from a finite execution trace.  These likely invariants can then be used to help formulate a compact model of a process which would generate such an event trace~\cite{RecoveringInvariants}.

\section{Conclusions}

We have shown a method of checking liveness and safety properties of CARMA models in which some components of the system which is being modelled are represented without abstraction, using concrete components which are generated automatically from data.  The properties which can be checked are those which can be expressed by finite-state automata (``probes'') whose state-to-state transitions are guarded by predicates over the values of component attributes or parameters passed by communication actions.  Checking is automatically performed by the CARMA Eclipse Plugin which can generate both graphs of the transitions of the probes and graphs of the changes in underlying values within the model (such as component attributes).  Using this we were able to detect from an unlabelled set of trajectories the trajectory which failed to satisfy the requirements of a specified bus route.  The methods used are generally applicable to any problem where data plays a significant role and whose correctness criterion can be expressed automata-theoretically.

Our interests for future work on this topic include generating probe components directly from route descriptions which list the bus stops on the route together with their latitude and longitude coordinates.  Additionally, we wish to add \emph{instrumentation} to probes in order that they can count visits to the regions of interest on the route, enabling stronger liveness properties to be expressed.

\iffalse
\section{Bibliography}

We request that you use
\href{http://www.cse.unsw.edu.au/~rvg/EPTCS/eptcs.bst}
{\tt $\backslash$bibliographystyle$\{$eptcs$\}$}. Compared to the original {\LaTeX}
{\tt $\backslash$biblio\-graphystyle$\{$plain$\}$},
it ignores the field {\tt month}, and uses the extra
bibtex fields {\tt eid}, {\tt doi}, {\tt ee} and {\tt url}.
The first is for electronic identifiers (typically the number $n$
indicating the $n^{\rm th}$ paper in an issue) of papers in electronic
journals that do not use page numbers. The other three are to refer,
with life links, to electronic incarnations of the paper.

Almost all publishers use digital object identifiers (DOIs) as a
persistent way to locate electronic publications. Prefixing the DOI of
any paper with {\tt http://dx.doi.org/} yields a URI that resolves to the
current location (URL) of the response page\footnote{Nowadays, papers
  that are published electronically tend
  to have a \emph{response page} that lists the title, authors and
  abstract of the paper, and links to the actual manifestations of
  the paper (e.g.\ as {\tt dvi}- or {\tt pdf}-file). Sometimes
  publishers charge money to access the paper itself, but the response
  page is always freely available.}
of that paper. When the location of the response page changes (for
instance through a merge of publishers), the DOI of the paper remains
the same and (through an update by the publisher) the corresponding
URI will then resolve to the new location. For that reason a reference
ought to contain the DOI of a paper, with a life link to corresponding
URI, rather than a direct reference or link to the current URL of
publisher's response page. This is the r\^ole of the bibtex field {\tt doi}.
DOIs of papers can often be found through
\url{http://www.crossref.org/guestquery};\footnote{For papers that will appear
  in EPTCS and use \href{http://www.cse.unsw.edu.au/~rvg/EPTCS/eptcs.bst}
  {\tt $\backslash$bibliographystyle$\{$eptcs$\}$} there is no need to
  find DOIs on this website, as EPTCS will look them up for you
  automatically upon submission of a first version of your paper;
  these DOIs can then be incorporated in the final version, together
  with the remaining DOIs that need to found at DBLP or publisher's webpages.}
the second method {\it Search on article title}, only using the {\bf
surname} of the first-listed author, works best.  
Other places to find DOIs are DBLP and the response pages for cited
papers (maintained by their publishers).
{\bf EPTCS requires the inclusion of a DOI in each cited paper, when available.}

Often an official publication is only available against payment, but
as a courtesy to readers that do not wish to pay, the authors also
make the paper available free of charge at a repository such as
\url{arXiv.org}. In such a case it is recommended to also refer and
link to the URL of the response page of the paper in such a
repository.  This can be done using the bibtex fields {\tt ee} or {\tt
url}, which are treated as synonyms.  These fields should not be used
to duplicate information that is already provided through the DOI of
the paper.
You can find archival-quality URL's for most recently published papers
in DBLP---they are in the bibtex-field {\tt ee}. In fact, it is often
useful to check your references against DBLP records anyway, or just find
them there in the first place.

When using {\LaTeX} rather than {\tt pdflatex} to typeset your paper, by
default no linebreaking within long URLs is allowed. This leads often
to very ugly output, that moreover is different from the output
generated when using {\tt pdflatex}. This problem is repaired when
invoking \href{http://www.cse.unsw.edu.au/~rvg/EPTCS/breakurl.sty}
{\tt $\backslash$usepackage$\{$breakurl$\}$}: it allows linebreaking
within links and yield the same output as obtained by default with
{\tt pdflatex}. 
When invoking {\tt pdflatex}, the package {\tt breakurl} is ignored.
\else\fi

\paragraph{Acknowledgements:} This work is supported by the EU QUANTICOL project, 600708.  We acknowledge the assistance of Transport for Edinburgh in providing access to their data on bus movement in the city of Edinburgh.  Our thanks go to Natalia Zo\'n for her help in developing the CARMA model used in this paper.  Our thanks also go to the developers of the CARMA Eclipse Plugin for providing such a useful and robust analysis platform.  We are grateful to the anonymous reviewers of this paper for many helpful suggestions for improvement.

\bibliographystyle{eptcs}
\bibliography{main,pepa}
\end{document}